\documentclass[aps,nofootinbib,showpacs,showkeys]{revtex4}
\usepackage{graphicx,amsmath}
\begin{document}

\title{Pentaquark $\Theta^+$, constituent quark structures, 
and prediction of charmed $\Theta_c^0$ and bottomed $\Theta_b^+$.}
\author{Kingman Cheung}
\affiliation{Department of Physics and NCTS, National Tsing Hua University,
Hsinchu, Taiwan R.O.C.}
\date{\today}

\begin{abstract}
The newly observed $\Theta^+$ resonance is believed to be a pentaquark
with the constituent quarks $uudd\bar s$.  There are a few options for
the constituent quark structure. Some advocate
diquark-diquark-antiquark $(ud)$-$(ud)$-$\bar s$ while others favor
diquark-triquark $(ud)$-$(ud\bar s)$.  We use the color-spin hyperfine
interaction to examine the energy levels of these structures, and we
find that the diquark-diquark-antiquark structure is slight favored.
We proceed to write down the flavor triplet and antisextet of the
charmed or bottomed exotic baryons with internal $qqqq\bar Q$ quarks.
We also estimate the mass of $\Theta_c^0$ and $\Theta_b^+$.
\end{abstract}
\preprint{}
\maketitle

\section{Introduction}

The recent discovery of the $\Theta^+ (1540)$ resonance 
\cite{exp1,exp2,exp3,exp4} has revived an old interest in bound states with
more than 3 constituent quarks (see e.g. Refs. \cite{jaffe,casmir,lipkin0}.)
The resonance has been observed in the reaction $\gamma ^{\scriptstyle 12}
{\rm C} \to K^-  \Theta^+ \to K^- K^+ n$ by LEPS \cite{exp1}, in 
$K^+ {\rm Xe} \to {\rm Xe}' \Theta^+ \to {\rm Xe}' K^0 p$ by DIANA \cite{exp2},
in $\gamma d \to K^- p \Theta^+ \to K^- p K^+ n$ by CLAS \cite{exp3}, and 
in $\gamma p \to K_s \Theta^+ \to K_s K^+ n$ by SAPHIR \cite{exp4}.
The mass of the resonance is at around 1540 MeV with a width of order $20$ MeV
and an isospin $I=0$.  The spin-parity is $\frac{1}{2}^+$.
Such a narrow width can be explained by an isospin-violating decay.

The interpretation of the $\Theta^+$ has been made in the constituent
quark model \cite{wilczek,lipkin1,lipkin2} and in the Skyrmion or
chiral soliton models \cite{skymion}.  There are also other studies 
\cite{others} related to the newly discovered $\Theta^+$.
In this work, we concentrate on
the constituent quark model.  Since $\Theta^+$ has the internal quarks
$uudd\bar s$, there are various possible configurations for this complicated
system.  The naive $K$-$N$ molecular interpretation involves only a weak
van-der-Waal-like force between the $K$ and $N$.  In general, the color
triplet, sextet, and octet interactions are much more attractive than 
the color singlet bond.  In view of this, Jaffe and Wilczek (JW) \cite{wilczek}
interpreted the bound state as a diquark-diquark-antiquark.  Each diquark
pair is in the $\bar{\textbf{3}}_c$ representation of SU(3)$_c$, and therefore
the system is like $\bar{\textbf{3}}_c \times \bar{\textbf{3}}_c \times
\bar{\textbf{3}}_c$, similar to a normal antibaryon.  Of course, the spin of
each diquark pair is different from a normal quark.  Its spin $S=0$.  Thus,
the two diquark pairs combine in a $P$-wave orbital angular momentum
to form a state with $\textbf{3}_c$ in color, 
spin $S=0$, and $\bar{\textbf{6}}_f$ in flavor.  Then, combining with
the antiquark to form a flavor antidecuplet and octet, with spin $S=1/2$.
The $\Theta^+$ is at the top of the antidecuplet and has an isospin $I=0$.

On the other hand, Karliner and Lipkin (KL) \cite{lipkin1,lipkin2} interpreted
the bound state as a diquark-triquark $(ud)$-$(ud\bar s)$.  The first 
stand-alone $(ud)$ diquark pair is in a state of spin $S=0$, color 
$\bar{\textbf{3}}_c$ and flavor $\bar{\textbf{3}}_f$  while the second
$(ud)$ diquark pair inside the cluster $(ud\bar s)$ is in a state of
spin $S=1$, color $\textbf{6}_c$ and flavor $\bar{\textbf{3}}_f$.
The triquark cluster is then in a state of spin $S=1/2$, color
$\textbf{3}_c$ and flavor $\bar{\textbf{6}}_f$.  So the overall configuration
will give a color singlet, spin $S=1/2$, and a flavor octet and antidecuplet.
The $\Theta^+$ is at the top of the antidecuplet and thus has $I=0$.  
This internal configuration of KL is different somewhat from that of JW.
The differences are (i) both the diquark pairs have the same
configuration in JW while in LK they are asymmetric,
(ii) the order of combining: in JW the diquark pairs are first combined to 
form the diquark-diquark subsystem before combining with the antiquark while
in KL the second diquark pair first combines with the antiquark to form
the triquark cluster, then combine with the diquark cluster, and
(iii) the color-spin hyperfine interaction would be different (we shall 
explain next.)

The constituent quark model has been successful in describing the meson 
and baryon spectra, with the mesons in the flavor singlet and octet, and
the baryons in the flavor singlet, octet, and decuplet.  The chromo-magnetic
(color-spin) 
hyperfine interaction was shown to be the dominant mechanism in the
determination of the mass splittings in the $S$-wave $q\bar q$ mesonic and 
$qqq$ baryonic systems \cite{ruju}.  It was widely believed that the same
is true in 4-quark and 5-quark systems \cite{jaffe,casmir}.  

In this note,
we shall employ the color-spin hyperfine interaction to investigate the
hyperfine energy levels of various quark configurations. 
\footnote
{There is another approach using flavor-spin hyperfine interaction 
\cite{flavor-spin} to study the stability of various configurations.
}
We found that the picture of diquark-diquark-antiquark of JW \cite{wilczek}
will give the most favorable hyperfine interaction while the picture
of KL \cite{lipkin1,lipkin2} 
has a slightly higher hyperfine interaction, but it does not mean that
it is unstable.  The difference in hyperfine interaction is less than 100 MeV,
which is in the same order as the uncertainty in the estimation.
We shall also point out that the naive treatment of
KL that there is no color-spin hyperfine interaction between the diquark
and triquark clusters is {\it not} adequate.  If we took their assumption,
we found that the difference in hyperfine interaction between the 
configurations of diquark-diquark-antiquark and diquark-triquark would be
of order 200 MeV, which is too large compared with the uncertainty.
We shall also extend to the charmed and bottomed baryons with the 
replacement of $\bar s \to \bar c,\, \bar b$.  We give the estimate for
the mass of $\Theta_c^0$ and $\Theta_b^+$.

\section{Color-spin Hyperfine Interaction}

The color-spin hyperfine interaction \cite{ruju} was shown to be
dominant in the determination of the mass splittings in the $S$-wave 
mesons and baryons.  Extensions to more
complicated quark systems were also performed (see e.g. 
\cite{jaffe,casmir}).
The Hamiltonian describing the color-spin hyperfine splitting of 
a multi-quark system is given by \cite{jaffe}
\[
H_{hf} = -V \sum_{i>j} \, (\vec \lambda_i \cdot \vec \lambda_j )\, 
                          (\vec \sigma_i \cdot \vec \sigma_j ) \;,
\]
where $\vec \lambda$ and $\vec \sigma$ denote, respectively, the matrices 
for the color SU(3)$_c$ and the spin SU(2), and $i,j$ are the quark labels.
The color SU(3)$_c$ and the spin SU(2) of the quarks can be combined in a
SU(6) color-spin symmetry.  For example, the fundamental representation 
for a quark in SU(6) is $\textbf{6}=(\textbf{3},1/2)$, where the 
first label inside the parenthesis is the representation for the SU(3)$_c$ 
and the second label is the spin.  We use the following notation to 
denote a particular quark configuration 
\[
\left| D_6,\;\; D_{3c}, \;\; S,\;\; N,\;\; 
D_{3f}  \right. \rangle  \;,
\]
where $D_6, D_{3c}, D_{3f}$ are representations in SU(6) of color-spin,
in SU(3)$_c$, and in SU(3)$_f$, respectively, $S$ is the spin of the system, 
and $N$ is the total number of quarks or antiquarks in the configuration.
A quark will be denoted by
$\left| \textbf{6},\, \textbf{3}_c,\, 1/2,\, 1,\, \textbf{3}_f \rangle
\right .$.

The general expression for the color-spin hyperfine splitting in systems
with quarks and antiquarks is given by \cite{jaffe}
\begin{equation}
\label{basic}
V_{hf} = \frac{v}{2} \left [ \bar C(total) - 2 \bar C (Q)  -2 \bar C(\bar Q)
+ 16 N \right ] \;,
\end{equation}
where $\bar C(total)$ refers to the whole system, $\bar C(Q)$ refers to
the subsystem of quarks only while $\bar C(\bar Q)$ refers to the subsystem
of antiquarks only.
Here
\begin{equation}
\bar C = C_6 - C_3 - \frac{8}{3} S (S+1)
\end{equation}
where $C_6(D_6)$ and $C_3(D_3)$ 
denote the Casimir of the representation $D_6$ in SU(6) and of $D_3$ in 
SU(3)$_c$, respectively.  The constant $v$ can be determined using the
hyperfine splitting between $N$ and $\Delta$:
\begin{equation}
V_{hf}(\Delta) - V_{hf}(N) = M_\Delta - M_N = 16 v = 293.08 \;{\rm MeV} \;,
\end{equation}
where we have used the average mass of $p$ and $n$ for $M_N$ \cite{pdg}.

\subsection{Diquark-diquark-antiquark}
First, we look at the possible structures of a diquark with their hyperfine
splittings
\begin{eqnarray}
\left| \textbf{21},\, \bar{\textbf{3}}_c,\, 0,\, 2,\, \bar{\textbf{3}}_f 
\rangle
\right. &:& -8 v \nonumber \\
\left| \textbf{21},\, \textbf{6}_c,\, 1,\, 2,\, \bar{\textbf{3}}_f \rangle
\right. &:& -\frac{4}{3} v \nonumber \\
\left| \textbf{15},\, \bar{\textbf{3}}_c,\, 1,\, 2,\, \textbf{6}_f \rangle
\right. &:& \frac{8}{3} v  \nonumber \\
\left| \textbf{15},\, \textbf{6}_c,\, 0,\, 2,\, \textbf{6}_f \rangle
\right. &:& 4 v \nonumber \;,
\end{eqnarray}
where we have required the combined wavefunction to be antisymmetric under
the interchange of the two quarks.  It is obvious that the first diquark is the
most stable, followed by the second one.  We shall keep these two structures
in the following discussion.  

Next, we combine the diquark-diquark.  Based on the fact that the 
diquark-diquark must be in $\textbf{3}_c$ in order to give a color singlet
with the antiquark, there are two possibilities to combine the diquark-diquark:
\begin{eqnarray}
\left| \textbf{21},\, \bar{\textbf{3}}_c,\, 0,\, 2,\, \bar{\textbf{3}}_f 
\rangle \right. & \otimes& 
\left| \textbf{21},\, \bar{\textbf{3}}_c,\, 0,\, 2,\, \bar{\textbf{3}}_f 
\rangle \right. \supset 
\left| \textbf{210},\, \textbf{3}_c,\, 0,\, 4,\, \textbf{3}_f + 
 \bar{\textbf{6}}_f \rangle \right.  \label{state1} \\
\left| \textbf{21},\, \bar{\textbf{3}}_c,\, 0,\, 2,\, \bar{\textbf{3}}_f 
\rangle \right.  & \otimes &
\left| \textbf{21},\, \textbf{6}_c,\, 1,\, 2,\, \bar{\textbf{3}}_f \rangle
\right. \supset
\left| \textbf{210},\, \textbf{3}_c,\, 1,\, 4,\, \textbf{3}_f + 
  \bar{\textbf{6}}_f \rangle \right.  \;. \label{state2}
\end{eqnarray}
Note that in the second combination it is also possible to have 
the $\textbf{105}$ that contains $(\textbf{3}_c,1)$, but however 
with a smaller Casimir.  Also note that the flavor in the diquark-diquark
configuration can either be a $\textbf{3}_f$ or a $\bar{\textbf{6}}_f$,
which is antisymmetric and symmetric, respectively, under the interchange
of the diquark pair.  Since the diquark-diquark system is a totally symmetric
state, the $\bar{\textbf{6}}_f$ must be combined with a spatially 
antisymmetric  state (i.e. a $P$-wave), while the 
$\textbf{3}_f$ has to combine with a spatially symmetric state (i.e. a 
$S$-wave).  Jaffe and Wilczek \cite{wilczek} argue that the blocking
repulsion may raise the energy of the spatially symmetric states, and 
so the $P$-wave state is preferred.  We use Eq. (\ref{basic})
to evaluate the color-spin hyperfine splitting between these two
diquark-diquark states:
\begin{equation}
\label{16}
V \left( \left| \textbf{210},\, \textbf{3}_c,\, 0,\, 4,\, \textbf{3}_f + 
 \bar{\textbf{6}}_f \rangle \right. \right ) = -16 v ,\;\;\;\;
V \left( \left| \textbf{210},\, \textbf{3}_c,\, 1,\, 4,\, \textbf{3}_f + 
  \bar{\textbf{6}}_f \rangle \right. \right ) = - \frac{40}{3} v \;.
\end{equation}
The first diquark-diquark configuration is relatively more
stable, but however the second configuration is only slightly higher in 
hyperfine level.
The next step is to combine the diquark-diquark with the antiquark, using 
the diquark-diquark state in Eq. (\ref{state1}), we have  
\begin{equation}
\label{state3}
\left| \textbf{210},\, \textbf{3}_c,\, 0,\, 4,\, \bar{\textbf{6}}_f 
\rangle \right.  \otimes 
\left| \bar{\textbf{6}},\, \bar{\textbf{3}}_c,\, 1/2,\, 1,\, 
\bar{\textbf{3}}_f \rangle \right .
\supset 
\left| \textbf{70},\, \textbf{1}_c,\, 1/2,\, 5,\, \textbf{8}_f +
\overline{\textbf{10}}_f \rangle \right . \;,
\end{equation}
which has a hyperfine splitting of
\begin{equation}
\label{jw}
V_{hf}\left(\left| \textbf{70},\, \textbf{1}_c,\, 1/2,\, 5,\, \textbf{8}_f +
\overline{\textbf{10}}_f \rangle \right . \right )
= -40 v \;.
\end{equation} 
On the other hand, combining the diquark-diquark state in Eq. (\ref{state2}) 
with the antiquark we have
\begin{equation}
\label{state4}
\left| \textbf{210},\, \textbf{3}_c,\, 1,\, 4,\, \bar{\textbf{6}}_f 
\rangle \right.  \otimes
\left| \bar{\textbf{6}},\, \bar{\textbf{3}}_c,\, 1/2,\, 1,\, 
\bar{\textbf{3}}_f \rangle \right .
\supset 
\left| \textbf{70},\, \textbf{1}_c,\, 1/2,\, 5,\, \textbf{8}_f +
\overline{\textbf{10}}_f \rangle \right . \;,
\end{equation}
which has a hyperfine splitting of 
\begin{equation}
V_{hf}\left(\left| \textbf{70},\, \textbf{1}_c,\, 1/2,\, 5,\, \textbf{8}_f +
\overline{\textbf{10}}_f \rangle \right . \right ) = - \frac{104}{3} v \;.
\end{equation}
Note that the spin $S=3/2$ state has to go with the $\textbf{1134}$ of the 
SU(6), which would give a much less favorable configuration.  
Although the final configurations in Eqs. (\ref{state3}) and (\ref{state4})
are the same, one of the diquark pairs in Eq. (\ref{state4}) is 
different in the spin, which induces the difference in the final 
hyperfine energy levels. 
Thus, in the picture of Jaffe and Wilczek the most favorable configuration
is (i) both the diquark pairs are in 
$\left| \textbf{21},\, \bar{\textbf{3}}_c,\, 0,\, 2,\, \bar{\textbf{3}}_f 
\rangle \right. $,
(ii) the diquark pairs are in a $P$-wave state, and
(iii) combining with the antiquark to produce 
$\left| \textbf{70},\, \textbf{1}_c,\, 1/2,\, 5,\, \textbf{8}_f +
\overline{\textbf{10}}_f \rangle \right .$, which has a positive parity.

\subsection{Diquark-triquark}

Karliner and Lipkin \cite{lipkin1,lipkin2} suggested that the $\Theta^+$ has 
an internal $(ud)$-$(ud\bar s)$ quark structure, in which the $(ud)$ diquark 
is an $I=0$ color antitriplet and the $(ud\bar s)$ triquark is an $I=0$
color triplet, with a $P$-wave orbital angular momentum between the two
clusters.  They also assumed that the color-spin hyperfine interaction
only operates within each cluster, but is negligible between the two clusters.
However, we will show below that the hyperfine interaction energy will be
further minimized if we consider the hyperfine interaction between
the clusters.

The configuration of the stand-alone diquark is
$\left| \textbf{21},\, \bar{\textbf{3}}_c,\, 0,\, 2,\, \bar{\textbf{3}}_f 
\rangle \right.$, while the diquark inside the triquark system has a 
configuration
$\left| \textbf{21},\, \textbf{6}_c,\, 1,\, 2,\, \bar{\textbf{3}}_f 
\rangle \right.$.  
Combining with the antiquark $\left| \bar{\textbf{6}},\, \bar{\textbf{3}}_c,\,
 1/2,\, 1,\, \bar{\textbf{3}}_f  \rangle \right.$, the triquark $(ud\bar s)$ 
has a configuration
$\left| \textbf{6},\, \textbf{3}_c,\, 1/2,\, 3,\, \bar{\textbf{6}}_f 
\rangle \right.$.  Then they evaluate the hyperfine splitting as
\begin{equation}
V_{hf} = V_{hf} ({\rm diquark}) + V_{hf}({\rm triquark}) 
       = - 8 v - \frac{56}{3} v = - \frac{80}{3} v \;.
\end{equation}
With this value of $V_{hf}$ the diquark-triquark configuration is not 
as stable as the diquark-diquark-antiquark configuration of Eq. (\ref{state3}).

However, we can evaluate more carefully the hyperfine energy of the
diquark-triquark system, including the interaction between the clusters,
 with Eq. (\ref{basic}):
\begin{equation}
V_{hf} = \frac{v}{2} \left [ \bar C(ud{\scriptscriptstyle -}ud\bar s) - 
2 \bar C(ud{\scriptscriptstyle -}ud)
-2 \bar C (\bar s) + 80 \right ] \;,
\end{equation}
where
\begin{eqnarray}
\bar C (ud{\scriptscriptstyle -}ud\bar s) &=& \bar C( 
 \left| \textbf{70},\, \textbf{1}_c,\, 1/2,\, 5,\, 
\textbf{8}+\overline{\textbf{10}}_f  \rangle \right. ) = 64 \nonumber \\
\bar C (ud{\scriptscriptstyle -}ud) &=& \bar C (
 \left| \textbf{105}+\textbf{210},\, \textbf{3}_c,\, 1,\, 4,\, 
\bar{\textbf{6}}_f  \rangle \right. ) =  \frac{224}{3}\;{\rm or} \;
       \frac{272}{3} \nonumber \\
\bar C(\bar s) &=& \bar C (
 \left| \bar{\textbf{6}},\, \bar{\textbf{3}}_c,\, 1/2,\, 1,\, 
\bar{\textbf{3}}_f  \rangle \right. ) =  16\nonumber  \;.
\end{eqnarray}
Then 
\begin{equation}
\label{ll}
V_{hf} = -\frac{56}{3} v \;\; {\rm or} \;\; -\frac{104}{3} v \;,
\end{equation}
which depends on whether one takes $\textbf{105}$ or $\textbf{210}$, 
respectively, for the $ud$-$ud$ system.  It happens that the naive assumption 
that there is no hyperfine interaction between the two clusters
\cite{lipkin2} gives the average of the two hyperfine
splittings obtained in Eq. (\ref{ll}).
At this point, we can compare the hyperfine levels of the configurations
of JW and LK.  According to 
Eq. (\ref{jw}) and the more negative one in Eq. (\ref{ll}), the 
diquark-diquark-antiquark configuration is slight more
favorable than the diquark-triquark configuration.

\subsection{Mass of $\Theta^+$}

Overall, the most favorable configuration is the diquark-diquark-antiquark
picture, in which both diquark pairs are in 
color and flavor antitriplet, spin $S=0$, in a $P$-wave orbital angular 
momentum.  The diquark-diquark then combine with the antiquark to form
a spin $S=1/2$, color singlet, and flavor octet or antidecuplet.
However, the quark configuration that the diquark-diquark is in a spin $S=1$
state of Eq. (\ref{state4}) is just slightly higher in energy level. 
We should keep this state as well in the following discussion.
In fact, this configuration has the same hyperfine energy level as the 
diquark-triquark picture, the hyperfine energy of which is given 
by the smaller value in Eq. (\ref{ll}).

We shall next estimate the mass of $\Theta^+$, using these two favorable 
configurations.
We use the approach of Karliner and Lipkin \cite{lipkin1,lipkin2}.
We separate the total hyperfine interaction into two portions: one from the
$(ud)$-$(ud)$ and another one from the interaction with the antiquark.  Thus,
from Eq. (\ref{jw}) and Eq. (\ref{16}),
\begin{equation}
V_{hf} = -16 v - 24 \left(\frac{m_u}{m_Q} \right ) v
\end{equation}
where $m_Q$ is the mass of the antiquark inside the $\Theta^+$.  
Now we can evaluate the hyperfine interactions of a nucleon $N$ and a meson
$(q \bar Q)$:
\begin{eqnarray}
V_{hf}(N) &=& - 8v \\
V_{hf} (q\bar Q) &=& - 16 \left( \frac{m_u}{m_Q} \right ) v \;.
\end{eqnarray}
We then take the difference in the hyperfine splitting as the mass difference:
\begin{equation}
\label{final1}
V_{hf}( ud{\scriptscriptstyle -}ud{\scriptscriptstyle -}\bar Q) - 
V_{hf}(N) -V_{hf}(q\bar Q) = 
M_{ud{\scriptscriptstyle -}ud{\scriptscriptstyle -}\bar Q} - M_N - M_{q\bar Q}
 = - \frac{1}{2} \left( 1+ \frac{m_u}{m_Q}
\right ) ( M_\Delta - M_N ) \;.
\end{equation}
On the other hand, if we use the quark configuration that the diquark-diquark
is in a spin $S=1$ state, the above equation becomes
\begin{equation}
\label{final2}
V_{hf}( ud{\scriptscriptstyle -}ud{\scriptscriptstyle -}\bar Q) 
- V_{hf}(N) -V_{hf}(q\bar Q) = 
M_{ud{\scriptscriptstyle -}ud{\scriptscriptstyle -}\bar Q} - M_N - 
M_{q\bar Q} = - \frac{1}{3} \left( 1+ \frac{m_u}{m_Q}
\right ) ( M_\Delta - M_N ) \;.
\end{equation}
We also have to estimate the $P$-wave excitation energy of $\Theta^+$.  
Instead of using the $D_s$ system, here we employ the mass difference
between $\Lambda(\frac{1}{2}^+)$ and $\Lambda(\frac{1}{2}^-)$ \cite{pdg}:
\begin{equation}
\label{55}
\delta P_s = M_{\Lambda(\frac{1}{2}^-)} - M_{\Lambda(\frac{1}{2}^+)} = 
290.3 \;{\rm MeV} \;.
\end{equation}
The reason we used this mass difference as the $P$-wave excitation energy 
is that this is the closest known system to the $\Theta^+$
that both systems contain exactly one strange antiquark and the rest being
light $u,d$ quarks. 
The hyperfine splitting between $\Lambda(\frac{1}{2}^+)$ and 
$\Lambda(\frac{1}{2}^-)$ is zero.
Thus, the mass of $\Theta^+$ is estimated to be, with $m_u/m_s \simeq 2/3$ 
\cite{lipkin1}, 
\begin{equation}
\label{20}
M_{\Theta^+} = \left \{ 
\begin{tabular}{ll} 1481 \; {\rm MeV} \;\;\; & if using Eq. (\ref{final1}) \\
                    1562 \; {\rm MeV} \;\;\; & if using Eq. (\ref{final2}) \;.
\end{tabular}
                \right .
\end{equation}
The observed mass of $\Theta^+$ is closer to the second value.
Some comments are in order.  

(i) The $P$-wave excitation energy estimated
here is almost 100 MeV larger than that in Ref. \cite{lipkin2}.  It means that
there is an intrinsic uncertainty of order 100 MeV in the estimation.

(ii) Although the observed mass of $\Theta^+$ is closer to the second
favorable configuration, it does not mean that the most favorable configuration
is wrong.  There are perhaps some unknown nonperturbative effects
involved in the five-quark bound states that may affect the most favorable 
configuration and the second most favorable configuration. 
Also, the $P$-wave excitation energy may be different in these two 
configurations, perhaps due to some orbital-spin interactions, which we 
naively ignore.  

(iii) Nevertheless, we believe the assumption taken by Karliner and Lipkin 
\cite{lipkin2} that there is no hyperfine interaction between
clusters is somewhat inadequate.  We found that if we took their naive 
assumption the difference in hyperfine
interaction between the most favorable configuration and their configuration
is $v( 8 + 16/3 (m_u/m_s) ) \approx 200$ MeV, which is too large compared 
with the uncertainty.

(iv) Straightly speaking, the hyperfine interaction formula in 
Eq. (\ref{basic}) is only applicable to $S$-wave hadronic systems.  Here we 
have taken the assumption that the color-spin hyperfine interaction in 
$P$-wave hadronic systems is the same as in the $S$-wave systems.  When
we compared the hyperfine energy levels of the JW and KL configurations, 
they are in the same orbital angular momentum.

(v) There is also a possible mixing between these two configurations. Let us
denote the diquark-diquark-antiquark configuration by $|a \rangle$ and
diquark-triquark by $|b\rangle$.  Allowing a mixing between these two states
we write the hyperfine interactions as
\[
\left( \langle a| \;\; \langle b | \right ) \;\left( 
  \begin{array}{cc}
        H_a  &  \delta h \\
        \delta h & H_b \end{array} \right )\;
\left( \begin{array}{c}
   |a \rangle  \\
   |b \rangle   \end{array} \right ) \;,
\]
where $H_a, H_b$ denote the hyperfine interaction of the state 
$|a\rangle$ and $|b\rangle$, respectively, and $\delta h$ denotes the
mixing. 
We can diagonalize the states through a mixing angle $\theta_{\rm mix}$
\begin{eqnarray}
  |a \rangle &=& \cos \theta_{\rm mix} | 1 \rangle  + \sin\theta_{\rm mix}
                                       | 2 \rangle  \nonumber \\
  |b \rangle &=& -\sin \theta_{\rm mix} | 1 \rangle  + \cos\theta_{\rm mix}
                                       | 2 \rangle  \nonumber \\
  \tan 2 \theta_{\rm mix} &=& \frac{2 \delta h}{H_b - H_a} \nonumber
 \end{eqnarray}
Assuming $\delta h \ll H_b - H_a$ the mixing angle $\theta_{\rm mix} \approx
         \delta h/(H_b -H_a)$, and we obtain the eigen-masses 
\begin{eqnarray}
   m_{1} &=& H_a - \frac{(\delta h)^2}{H_b - H_a} \nonumber \\
   m_{2} &=& H_b + \frac{(\delta h)^2}{H_b - H_a} \nonumber 
\end{eqnarray}
Therefore, the mass splitting between $m_1$ and $m_2$ is
\begin{equation} 
m_2 - m_1 = H_b - H_a + \frac{2 (\delta h)^2}{H_b- H_a} \;,
\end{equation}
which implies that the splitting between the two configurations is increased
by a factor, which depends on the parameter $\delta h$.  Based on this 
mixing argument the mass estimates in Eq. (\ref{20}) will be modified 
such that the smaller one is lowered while the larger one is raised
by an amount $(\delta h)^2/( 81 {\rm MeV})$.
 
\section{Charmed Pentaquark $\Theta_c^0$}

\begin{figure}
\centering
\includegraphics[width=4in]{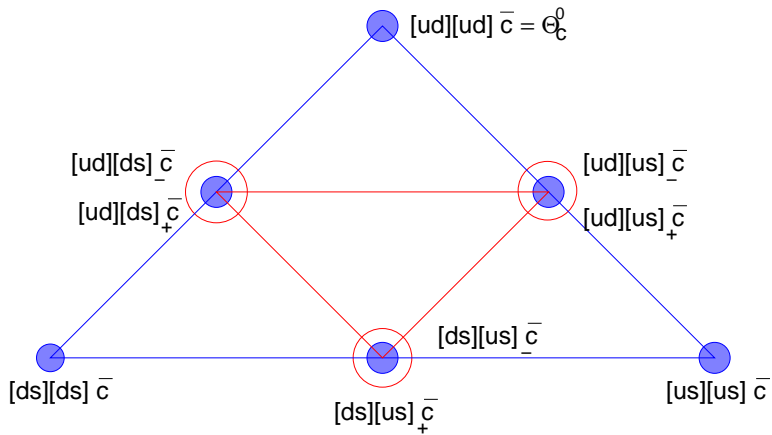}
\caption{\small \label{fig1}
The flavor triplet and antisextet of the charmed baryons with a charm antiquark
and 4 light quarks ($u,d,s$). The triplet consists of the three antisymmetric
pairs of $[ud][ds]_-$, $[ud][us]_-$, and $[ds][us]_-$, while those with 
$[ud][ds]_+$, $[ud][us]_+$, and $[ds][us]_+$
belong to the symmetric antisextet.
}
\end{figure}

The diquark-diquark-antiquark picture of Jaffe and Wilczek \cite{wilczek}
can be easily extended to charmed pentaquark, with the replacement
$\bar s \to \bar c$.  Since the charm quark does not belong to the
SU(3)$_f$ of $(u,d,s)$, the internal quark configuration of $\Theta_c^0$ will
follow the configuration of the diquark-diquark subsystem, which is
\begin{equation}
\left| \textbf{210},\, \textbf{3}_c,\, 0,\, 4,\, \textbf{3}_f +
\bar{\textbf{6}}_f  \rangle \right. \;.
\end{equation}
For the $\textbf{3}_f$ the diquark-diquark will be in a $S$-wave state
while for $\bar{\textbf{6}}_f$ it will be in a $P$-wave state.
The flavor triplet and antisextet are shown in Fig. \ref{fig1}.
Pentaquark baryons in the framework of Skyrme model were considered before
\cite{oh}.
There also exist upper limits on the production of the isodoublet 
$|\bar csuud \rangle$ and $|\bar csddu \rangle$ reported 
by the E791 collaboration \cite{E791}.

We can also estimate the mass of $\Theta_c^0 \equiv (ud)$-$(ud)$-$\bar c$.  
\footnote{
Jaffe and Wilczek \cite{wilczek} also estimated the mass of $\Theta_c^0$
and $\Theta_b^+$.  They used 
$M_{\Theta_c^0} - M_{\Theta^+} = M_{\Lambda_c} - M_{\Lambda}=1170$ MeV, and
similarly for $\Theta_b^+$.  They obtained substantially lower masses 
than our estimates.  Our estimates are close to those by Karliner and 
Lipkin \cite{lipkin2}.  
}
The formula is similar to the one for $\Theta^+$ 
(analogous to Eq.(\ref{final2}))
\begin{equation}
M_{\Theta_c^0} = M_N + M_D  - \left( \begin{array}{c}
                                      1/2 \\
                                      1/3 \end{array} \right )
\left(1 + \frac{m_u}{m_c} \right )  (M_\Delta - M_N )  
+ \delta P_c \;,
\end{equation}
where the $P$-wave excitation energy is estimated by the mass difference
between $\Lambda_c (\frac{1}{2}^+)$ and $\Lambda_c (\frac{1}{2}^-)$ \cite{pdg}
such that $\delta P_c = 309$ MeV.
\footnote{This is analogous to what we used in Eq. (\ref{55}), the
$\Lambda_c(\frac{1}{2}^+)$ and $\Lambda_c(\frac{1}{2}^-)$ system contains
exactly one charm antiquark and the rest being $u,d$ light quarks.}
  The mass of $\Theta_c^0$ is then given by,
with $m_u/m_c \approx 0.21$ \cite{lipkin1},
\begin{equation}
M_{\Theta_c^0} = \left \{ 
\begin{array}{c} 2938 \; {\rm MeV} \\
                 2997 \; {\rm MeV} 
\end{array}
                \right . \;.
\end{equation}
Again, the major uncertainty comes from the estimation of the $P$-wave
excitation energy, which is of order of 100 MeV.
Similarly, we can estimate the bottomed baryon $\Theta_b^+$ using
\begin{equation}
M_{\Theta_b^+} = M_N + M_B  - \left( \begin{array}{c}
                                      1/2 \\
                                      1/3 \end{array} \right )
\left(1 + \frac{m_u}{m_b} \right )  (M_\Delta - M_N )  
+ \delta P_b \;.
\end{equation}
Since the $\Lambda_b(\frac{1}{2}^-)$ has not been found experimentally 
\cite{pdg}, 
we use $\delta P_b \approx \delta P_c$, which is reasonable because
$\delta P_c \approx \delta P_s$.  Therefore, we obtain with $m_u/m_b \approx
0.071$ \cite{lipkin1}
\begin{equation}
M_{\Theta_b^+} = \left \{ 
\begin{array}{c} 6370  \; {\rm MeV} \\
                 6422  \; {\rm MeV} 
\end{array}
                \right . \;.
\end{equation}

\section{Conclusions}

We have used the color-spin hyperfine interaction to examine the
hyperfine energy levels of various quark configurations. 
We found that the picture of diquark-diquark-antiquark of Jaffe and 
Wilczek \cite{wilczek}
gives the most favorable hyperfine interaction while the picture
of Karliner and Lipkin \cite{lipkin1,lipkin2} 
has a slightly higher hyperfine interaction, but it does not mean that
it is unstable.  
The observed mass of $\Theta^+$ is in between
the two hyperfine levels of our estimation, but the uncertainty in the 
estimation is of order of 100 MeV.
We have also predicted a flavor triplet and antisextet for the charmed and
bottomed pentaquark baryons.  The mass of the $\Theta_c^0$ is between
2938 and 2997 MeV whereas the mass of $\Theta_b^+$ is between
6370 and 6422 MeV.

The work was supported by the NSC, Taiwan.  We thank Robert Jaffe for a 
correspondence.

\end{document}